\documentclass[9pt,conference]{IEEEtran}

\IEEEoverridecommandlockouts
\usepackage{multirow}
\usepackage{adjustbox}
\usepackage{caption}
\usepackage{subcaption}
\usepackage{stackengine}
\usepackage{amsmath}
\usepackage{hyperref}
\usepackage{booktabs}

\usepackage{csquotes}
\usepackage{cite}
\usepackage{amsmath,amssymb,amsfonts}
\usepackage{algorithmic}
\usepackage{graphicx}
\usepackage{textcomp}
\usepackage{xcolor}
\def\BibTeX{{\rm B\kern-.05em{\sc i\kern-.025em b}\kern-.08em
    T\kern-.1667em\lower.7ex\hbox{E}\kern-.125emX}}

\newcommand{\etal}{\textit{et al.}}

\begin{document}

\title{Speaking Without Sound: Multi-speaker Silent Speech Voicing with Facial Inputs Only}

\author{
\IEEEauthorblockN{
Jaejun Lee$^{\star}$,
Yoori Oh$^{\star}$,
Kyogu Lee$^{\star\dagger\ddagger}$\\
\thanks{This paper was presented at ICASSP 2025.}
    }
\IEEEauthorblockA{
    $^{\star}$Music and Audio Research Group (MARG),
    Department of Intelligence and Information,\\
    $^{\dagger}$Interdisciplinary Program in Artificial Intelligence,
    $^{\ddagger}$Artificial Intelligence Institute, Seoul National University\\
    \{jjlee0721, yoori0203, kglee\}@snu.ac.kr\\
    }
}
\maketitle

\begin{abstract}
In this paper, we introduce a novel framework for generating multi-speaker speech without relying on any audible inputs.
Our approach leverages silent electromyography (EMG) signals to capture linguistic content, while facial images are used to match with the vocal identity of the target speaker.
Notably, we present a pitch-disentangled content embedding that enhances the extraction of linguistic content from EMG signals.
Extensive analysis demonstrates that our method can generate multi-speaker speech without any audible inputs and confirms the effectiveness of the proposed pitch-disentanglement approach.
\end{abstract}

\begin{IEEEkeywords}
Electromyography, Silent speech interface, voice conversion, cross-modal generation
\end{IEEEkeywords}

\section{Introduction}
Speech is a fundamental aspect of social communication,
and individuals with speech impairments often face challenges beyond the inability to speak, including social isolation and diminished self-esteem \cite{wadman2008self}. 
In recent years, biosignal-based spoken communication technologies, such as Silent-Speech Interfaces (SSI) \cite{denby2010silent}, have seen rapid advancements, offering new possibilities for those affected by speech impairments \cite{gonzalez2020silent}.

% Electromyography (EMG) captures articulatory behavior during speech by detecting signals from electrodes placed on facial muscles. Notably, recent advancements have focused on surface EMG, which is non-invasive, making it a more practical approach \cite{gaddy2020digital, scheck2023multi}. This progress suggests that if an individual can perform the articulatory movements, whether audible or silent, it is possible to generate voiced speech using these techniques.

Electromyography (EMG) captures articulatory behavior during speech by detecting signals from electrodes placed on facial muscles.
Notably, Recent work has shown that speech can be generated from EMG signals without requiring audible input~\cite{gaddy2020digital, scheck2023multi}.
% Notably, recent work has shown promising results in generating speech solely from EMG signals, meaning that no audible input is required to convey speech content \cite{gaddy2020digital, scheck2023multi}. 
However, in a multi-speaker setting, this approach still relies on the availability of the target speaker's audible speech to imprint their speaking style. This limitation suggests that individuals with speech impairments, who are unable to produce audible speech, may not fully benefit from this type of technology.

Recent advancements in face-based voice conversion have successfully associated target voice characteristics with facial images, allowing the target speaker's timbre to be imprinted onto source speech audio \cite{sheng2023face, lee24d_interspeech}. This means that audible speech is not required to transfer the target speaker's timbre, as long as a facial image of the target is available.

% In this work, we present a multi-speaker silent speech generation framework that uses only facial inputs, silent EMG for linguistic contents and facial image for timbre. The contribution we claim is as follows,
% \begin{itemize}
% \item We present a multi-speaker speech generation framework that never dealing with before, which doesn't need any audible inputs.
% \item A novel approach pitch-disentangled content embedding, which is effective on EMG based speech generation, is proposed.
% \item Through extensive analysis, we demonstrate our method can generate not only have naturalness of its speech quality, and also shows its timbre aligns with target speaker's voice characteristics.
% \end{itemize}
In this work, we introduce a multi-speaker silent speech generation framework that relies solely on facial inputs, utilizing silent EMG signals for linguistic content and facial images for timbre. The key contributions of our work are as follows:
\begin{itemize}
\item We present a novel multi-speaker speech generation framework that operates without any audible inputs, a concept not previously explored.
\item We propose a new approach, pitch-disentangled content embedding, which proves effective in EMG-based speech generation.
% \item Through extensive analysis and our newly proposed evaluation metric, we demonstrate that our method not only produces speech with natural quality but also successfully aligns the generated speech's timbre with the target speaker's voice characteristics.
\item Through extensive analysis and our newly proposed evaluation metric, we demonstrate that our method not only generates audible speech from silent EMG but also aligns it with the target speaker's voice characteristics using only facial images.
% Additionally, it better estimates the pitch contour of the source content.
\end{itemize}
The demo is available on, \url{https://jaejunL.github.io/SWS_Demo/}.

\section{Related work}
\subsection{EMG-to-speech}
Electromyography (EMG) detects electric signals from electrodes placed on facial muscles to capture articulation during speech. A key advantage is that EMG captures signals whether the speech is audible or silent, and it detects the electric signal 60ms before actual articulatory movements occur \cite{jou2006towards}.
A new EMG dataset has been introduced, containing both voiced EMG (collected during voiced speech) and silent EMG (collected during silent speech)
 from a single speaker~\cite{gaddy2020digital}. Since parallel speech data is available only for voiced EMG, the authors propose a training approach using Dynamic Time Warping (DTW) loss to align silent EMG with voiced speech. By employing this approach along with a transformer-based EMG~encoder, they have demonstrated significant results in generating speech from silent EMG.
However, as this work is limited to single-speaker speech generation, Sheck \etal~\cite{scheck2023multi} have proposed a multi-speaker EMG-to-speech generation framework. This framework utilizes voice conversion \cite{van2022comparison} by disentangling content embedding and target speaker embedding to imprint the style of an unseen target speaker.
Although this work enables multi-speaker settings, it still requires audio data from the target speaker, which may not be available in certain scenarios, such as cases involving speech impairments.
Additionally, previous research has directly mapped EMG signals to speech features, which can lead to information mismatches, such as pitch, which is present in speech but not in EMG signals particularly in silent~EMG.

\begin{figure*}[t!]
\centerline{\includegraphics[width=2.08\columnwidth]{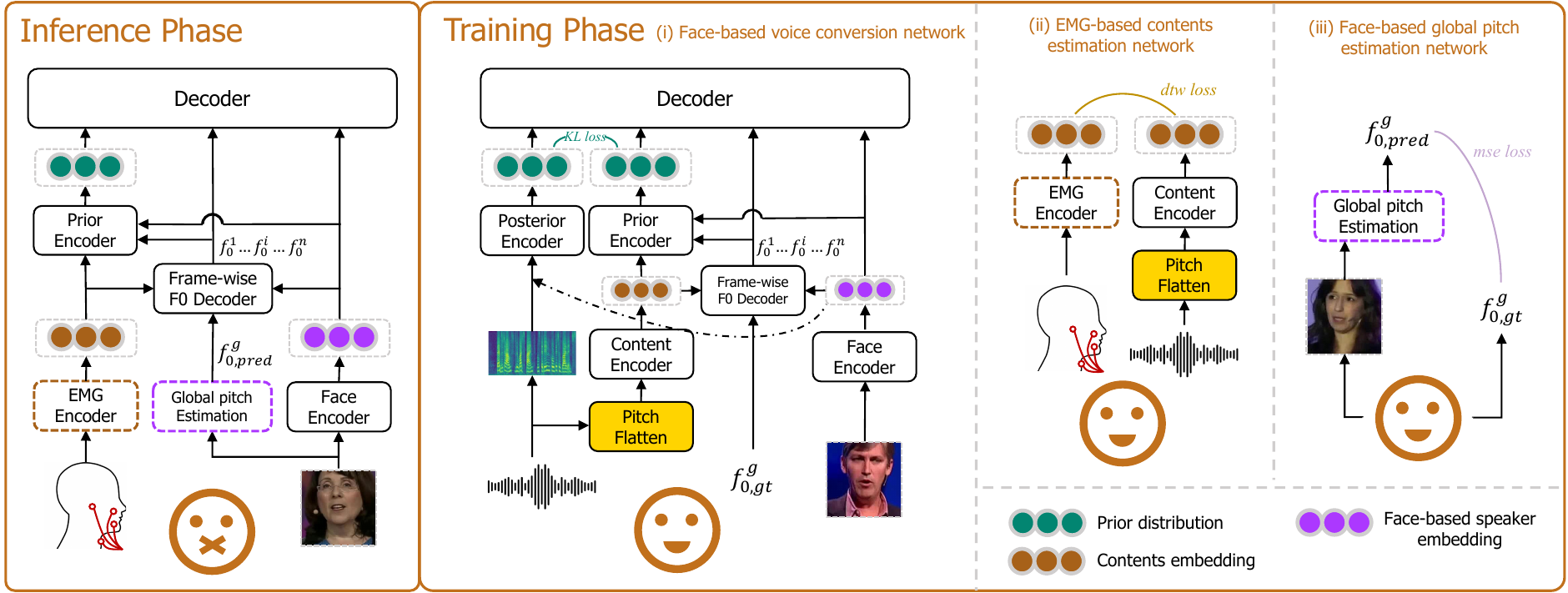}}
\centerline{\hspace{0.4cm}(a) Inference phase\hspace{6.2cm} (b) Training phase\hspace{5cm}}
\caption{Overview of the proposed multi-speaker EMG-to-Speech generation framework. In the (a) Inference phase, content embedding is estimated from EMG signals, while speaker embedding and global pitch information are derived from a facial image.
% These representations are then used to predict the prior distribution of speech with frame-wise $F0$ information.
During the (b) Training phase, for training the (i)~face-based voice conversion network, a predefined speaker-wise global pitch ($f_{0,\mathit{gt}}^{\mathit{g}}$) is used to estimate frame-wise $F0$ values. However, since the $f_{0,\mathit{gt}}^{\mathit{g}}$ values for unseen target speakers are not available during the inference, a (iii)~face-based global pitch estimation network is independently trained using only face images.
Additionally, as the speech-based content encoder is not available during inference, the (ii)~EMG-based contents estimation network is trained.
% This module is then utilized in the inference phase.
Each network is trained independently during the training phase.}
\label{fig:figure1}
\end{figure*}

\subsection{Face-based voice conversion}
Recent studies reveal that human have ability to match facial images with the voice characteristics of the same speaker \cite{kim2019learning, campanella2007integrating, mavica2013matching}. Building on these findings, face-based speech synthesis has surged and demonstrated prominent results \cite{goto2020face2speech, lu2021face, sheng2023face}.
Our previous work~\cite{lee24d_interspeech} demonstrated that using a face-based voice conversion framework, it is possible to generate speech that reflects the voice characteristics of a target speaker using only a facial image, even when audible information from the target speaker is unavailable. The method not only estimates the speaker embedding from the facial image but also predicts the speaker's global pitch.
However, these approaches still rely on `\textit{voice}' conversion, which requires source audio that contains the content information.

\section{Methods} \label{sec:method}
\subsection{Multi-speaker EMG-to-Speech generation} \label{sec:method1}
Our proposed method is built on a face-based voice conversion framework with a conditional variational autoencoder architecture, similar to \cite{lee24d_interspeech}.
Unlike the conventional face-based voice conversion networks that use source audio to extract content information, we employ an EMG-based content estimation network.
% similar to \cite{gaddy2020digital, scheck2023multi},
 % to derive content information from EMG signals.
We also utilize face-based global pitch estimation, as in \cite{lee24d_interspeech}, to more effectively model target voice characteristics from facial images.
Additionally we propose a pitch-flattening module to enhance EMG-based content estimation.
Note that all modules are trained from scratch, except for the content encoder, which uses pretrained SSL representations.
Figure~\ref{fig:figure1} illustrates the entire framework of our proposed method, including (a)~Inference phase and (b)~Training phase.
A detailed explanation of each module is as follows.

\subsubsection{Face-based voice conversion network} \label{sec:method1-1}
This is a conventional voice conversion framework where the facial image of the target speaker is used to modify the speaker information of the source speech. A content embedding ($c$) is fed into the prior encoder, while the face-based speaker embedding ($s$) conditions both the prior encoder and the posterior encoder.
Additionally, similar to \cite{lee24d_interspeech}, frame-wise $F0$ values, $f_0^i$ ($i=1,...,n$. $n$ is the number of frames) conditions both the prior encoder and the decoder.
A speaker-wise global pitch~($f_{0,\mathit{gt}}^{\mathit{g}}$) 
is adjusted to $f_0^i$ within the the frame-wise $F0$ decoder ($\mathit{FF}$), in conjunction with the $c$ and $s$.
The loss~$\mathcal{L}_{\mathit{ff}}$ for training $\mathit{FF}$ is as follows:
\vspace{-2mm}
\begin{equation}
\mathcal{L}_{\mathit{ff}}=\frac{1}{n}\sum_{i=1}^n(f_{0,\mathit{gt}}^i-\mathit{FF}(f_{0,\mathit{gt}}^{\mathit{g}}, c, s))^2,
\vspace{-2mm}
\end{equation}
where $f_{0,\mathit{gt}}^i$ refers to the ground-truth frame-wise $F0$ values.
Note that the global pitch, $f_{0,\mathit{gt}}^{\mathit{g}}$ value represents the average of $f_{0,\mathit{gt}}^i$ values across all speech frames for each speaker in the training dataset.
Not only the loss~$\mathcal{L}_{\mathit{ff}}$, we used the mel-spectrogram based reconstruction loss~\cite{kim2021conditional}, KL~(Kullback-Leibler) divergence loss for the prior encoder, adversarial loss~\cite{mao2017least}, and feature matching loss~\cite{larsen2016autoencoding} for the generation quality.

\subsubsection{EMG-based contents estimation} \label{sec:method1-2}

While the face-based voice conversion network relies on speech based content embedding $c$, a separate network is trained to generate content embedding from EMG signals ($c_{emg}$), similar to the approach in \cite{gaddy2020digital}. Voiced EMG 
% (EMG signals collected during voiced speech)
naturally pairs with speech audio of the same length, making it straightforward to train EMG signals to produce $c$ of the paired speech. However, silent EMG 
% (EMG signals collected during silent speech)
lacks paired speech audio.
To address this, Gaddy~\cite{gaddy2020digital} created a dataset where both voiced and silent EMG share the same content. This means that the speaker utters the same phrase twice—once vocally and once silently.
However, even when paired speech audio for silent EMG is available, the lengths often differ. To address this, the author used the dynamic time warping (DTW) algorithm to align $c_{emg}$ with the $c$ of the voiced speech. Consequently, the loss function we use is defined as follows:
\begin{equation}
\mathcal{L}_{\mathit{c}}=||\mathbf{c}-DTW(\mathbf{c}_{emg})||_2,
\end{equation}
Similarly, as in \cite{gaddy2020digital}, we also used a Cross-Entropy based phoneme classification loss, as the dataset includes phoneme label information.

\subsubsection{Face-based global pitch estimation} \label{sec:method1-3}
In the absence of global pitch ($f_{0,\mathit{gt}}^{\mathit{g}}$) information during the inference phase, we independently train a face-based global pitch estimation network ($\mathit{GP}$) using only the facial images ($v$) of target speakers as in~\cite{lee24d_interspeech}. The loss~$\mathcal{L}_{\mathit{gp}}$ function is as follows:
\begin{equation}
\mathcal{L}_{\mathit{gp}}=(f_{0,\mathit{gt}}^{\mathit{g}}-\mathit{GP}(v))^2.
\end{equation}
Note that, $f_{0,\mathit{gt}}^{\mathit{g}}$ is a scalar value for each speaker, representing their characteristic general pitch information independent of content information.

\subsubsection{Pitch-flattening}   \label{sec:method1-4}
Given the inherent nature of silent EMG signals, which capture only the articulatory movements of facial muscles, the information gathered is expected to primarily reflect linguistic content and duration, rather than any pitch information in speech. However, the ground-truth content embedding might still contain pitch information, as it is derived from speech data.

To address this, we propose a pitch-flattening module that removes pitch variance from the ground-truth speech data. It is a classical method based on Praat~\cite{boersma2001speak}, specifically the `pitch manipulation' technique. It uses auto-correlation method~\cite{boersma1993accurate} to analyze pitch information and the overlap-add resynthesis method, commonly known as PSOLA~\cite{moulines1990pitch}, to create flattened-pitch speech.

Unlike the global pitch estimation, which treats pitch as a characteristic of the target speaker's voice, the pitch-flattening module is designed to remove content-related pitch information from $c$. If the face-based voice conversion network uses $c$ without pitch-flattening, the frame-wise $F0$ decoder ($FF$) will extract pitch information directly from $c$ rather than estimating it using both content and speaker information.
Consequently, during the inference phase, the mismatch in pitch information within the EMG-based content embedding $c_{emg}$ could lead to performance degradation, as $c_{emg}$ naturally does not contain pitch information, especially for the silent EMG cases.
The proposed pitch-flattening module forces the $FF$ to work harder by estimating content-related pitch information using $c$, along with global pitch and speaker embedding, rather than simply extracting it from $c$. This module is applied before the content encoder in both the face-based conversion network and the EMG-based content estimation network.

\subsection{Model architecture}
\noindent\textbf{Posterior Encoder}: WaveNet-based residual blocks, conditioned by face-based speaker embeddings, with a method similar to~\cite{kim2021conditional}.\\
\textbf{Prior Encoder}: Transformer-based architecture similar to \cite{shaw2018self}, atop which is stacked a normalizing flow layer
% comprised of residual coupling blocks
\cite{dinh2016density}.\\
\textbf{Face Encoder}: Vision Transformer \cite{dosovitskiy2020image} with projection layer.\\
\textbf{Contents Encoder}: ContentVec, a pretrained SSL representation~\cite{qian2022contentvec}, specifically using the hugging face version\footnote{https://huggingface.co/lengyue233/content-vec-best}\\
\textbf{Decoder}: Similar to the generator of HiFi-GAN \cite{kong2020hifi} so as to discriminator network ($D$). However, for more precise conditioning of $F0$ information, a neural source-filter method \cite{wang2019neural} based conditioning similar to Sovits-SVC\footnote{https://github.com/svc-develop-team/so-vits-svc} is employed.\\
\textbf{Frame-wise $\textbf{F0}$ Decoder ($\textbf{FF}$}): Self-attention layers and feed forward layers conditioned with both content embedding and face-based speaker embedding. Fast Context-base Pitch Estimator\footnote{https://github.com/CNChTu/FCPE} (FCPE) is used to extract pitch information, including frame-wise $F0$ values and speaker-wise global pitch values ($f_{0,\mathit{gt}}^{\mathit{g}}$), as well as voice activity information to identify non-speech segments.\\
\textbf{EMG encoder}: Transformer encoder \cite{vaswani2017attention}, similar to 
% those used in 
\cite{gaddy2020digital, scheck2023multi}.\\
\textbf{Global pitch estimation network (GP)}: Similar to face encoder, vision transformer based architectures with projection layer.\\
\textbf{Pitch flatten}: Parselmouth interface \cite{jadoul2018introducing} of Praat \cite{boersma2001speak}\\

\section{Experiments}
\subsection{Dataset}
\subsubsection{EMG dataset}
As depicted in Section~\ref{sec:method1-2}, to train the EMG-based contents estimation network, we used Gaddy's~\cite{gaddy2020digital} dataset, which contains approximately 20 hours of single-speaker EMG signals. This dataset 
 includes both voiced and silent EMG signals, with paired speech available only for the voiced EMG. We followed the predefined split for training, validation, and testing, and adhered to all preprocessing steps, including filtering and normalization~\footnote{https://github.com/dgaddy/silent\_speech}.

\subsubsection{Speech and facial image dataset}
To Train the face-based voice conversion network~(Section~\ref{sec:method1-1}) and the face-based global pitch estimation network~(Section~\ref{sec:method1-3}), we used the LRS3 \cite{afouras2018lrs3} dataset, which consists of 5,502 TED and TEDx videos totaling over 430 hours. Each video includes 224x224 resolution image at 25 frames per second, with 16kHz audio. We followed the predefined splits and used only the center-cropped image of size 112x112, as in \cite{lee24d_interspeech}.

\subsubsection{Evaluation set}
Given our goal of multi-speaker silent EMG voicing, we selected 50 male and 50 female speakers from the testing split of the LRS3 dataset, similar to \cite{lee24d_interspeech}. We randomly choose 2~speakers from each gender group for each test sample, and all experiments were repeated 10 times. This resulted in a test set size that is 40 times ($2*n_{test}*2(genders)*10(trials)$) larger for the original test set ($n_{test}$). The results were averaged, and we reported the outcomes for both the voiced and silent EMG test sets.

\subsection{Metrics}
\subsubsection{Intelligibility}
For evaluating speech intelligibility, we reported overall ASR error rates, including Word Error Rates (WER) and Character Error Rates (CER). Similar to \cite{scheck2023multi}, we used the Whisper~\cite{radford2023robust}, ``\textit{medium.en}'' model.

\subsubsection{Consistency}
For evaluating speaker consistency, we assessed both objective and subjective consistency. Objective consistency involves comparing the speaker embedding cosine similarity of the synthesized audio with that of the ground-truth audio from the same speaker. To ensure a robust comparison, we also evaluated this metric with ground-truth audio from a random speaker. Speaker embeddings were generated using \textit{Resemblyzer}\footnote{https://github.com/resemble-ai/Resemblyzer}.
Subjective consistency was measured through subjective evaluation using a 5-point MOS scale (completely inconsistent to completely consistent) to determine whether the synthesized audio aligns with the corresponding facial images. We followed the same procedure as in \cite{lee24d_interspeech}, utilizing Amzon Mechanical Turk (MTurk).

\begin{table}[t!]
\caption{Intelligibility evaluation.}
\centering
\begin{adjustbox}{width=0.35\textwidth}
\begin{tabular}{ccccc}
\hline\hline
\multirow{2}{*}{} & \multicolumn{2}{c}{WER$\downarrow$} & \multicolumn{2}{c}{CER$\downarrow$}\\
% & \multicolumn{2}{c}{Const(obj)$\uparrow$}\\
\cmidrule(lr){2-3} \cmidrule(lr){4-5}
 & Voiced & Silent & Voiced & Silent\\
 % & Voiced & Silent\\
\hline
GT & 2.85 & - & 0.90 & - \\
\hline
Base & \textbf{16.98} & 40.12 & \textbf{9.96} & 26.14\\
Flatten & 17.48 & \textbf{38.91} & 13.28 & \textbf{24.28}\\
\hline\hline
\end{tabular}
\end{adjustbox}
\label{table:result1}
\vspace{-2mm}
\end{table}

\subsubsection{Pitch evaluation}
We evaluate two metrics related to pitch. The first is \textit{Global F0}, as proposed in \cite{lee24d_interspeech}, which assesses the model's ability to track the voice characteristics of the target speaker.
The second is our newly proposed metric, called \textit{Local F0 deviations}. It is crucial to assess whether the model can generate natural, content-related pitch information, given that silent EMG signals inherently do not contain pitch information. To address this, we compare the frame-wise pitch deviation of the synthesized audio ($f_{0,pred}^{i,dv}$) with that of the ground-truth audio (speech with the same content) ($f_{0,gt}^{i,dv}$).
% , thereby assessing the model's ability to generate content-related pitch using EMG signals.
The term `deviation (\textit{dv})' here is used because we normalize the global pitch, allowing us to isolate and analyze only the content-related pitch deviations, as the error in global pitch estimation is usually much larger.
Therefore, the proposed metric \textit{Local F0 deviations} is defined as
% frame-wise pitch deviation $f_{0_{dv}}^i$ ($i=1,...,n$. $n$ is the number of frames) is defined as,
% \vspace{-2mm}
\begin{equation} \label{eq:1}
||f_{0,{pred}}^{i,dv}-f_{0,{gt}}^{i,dv}||_2
\end{equation}

where $f_{0,x}^{i,dv} = f_{0,x}^i-{\sum_{i=1}^n(f_{0,x}^i*u_x^i)}/\sum_{i=1}^n(u_x^i)$,~$x\in~\{pred, gt\}$, 
$n$ is the number of frames and $u^i$ is the output of the voice activity detection, indicating whether the frame contains voice (1) or non-voice (0).
% The right-hand term of Equation~\ref{eq:1} represents the averaged pitch of the input audio, 
% ensuring that the $f_{0_{dv}}^i$ value reflects only the pitch deviation, independent of the speaker's global pitch.
% The metric is therefore, defined as $f_{0_{dv}}^{i,{pred}}-f_{0_{dv}}^{i,g}$.
To evaluate \textit{Local F0 deviations} it is necessary to synthesize speech that has the same content as the ground-truth speech-in Gaddy's dataset. Therefore, we used the facial image from the author's paper~\cite{gaddy2020digital} as well as EMG signal data.
% Unlike conventional voice conversion,
% It is crucial to assess whether the model can generate natural, content-related pitch information, given that silent EMG signals inherently do not contain pitch information.
We note that, for evaluating the speech synthesized from silent EMG, we use the speech paired with voiced EMG as the ground truth. Although the content is the same, the lengths often differ, so we apply nearest interpolation to align the frame lengths in the silent case.

% We also evaluated the \textit{Global F0}, as proposed in \cite{lee24d_interspeech}, which compares the deviations between the $f_0^{avg}$ of the synthesized audio and the global pitch (represented as $f_{0,\mathit{gt}}^{\mathit{g}}$ in Section~\ref{sec:method}) of the ground-truth target speaker. This evaluation assesses the model's ability to capture explicit voice characteristic of the target speaker.

% Note that the standard deviations (stdv) of the $F0$ for all audio samples are 29.18 Hz for male speakers and 37.50 Hz for female speakers. These values serve as a baseline, reflecting the deviation is based solely on gender class. If the model captures the associations between facial features and voice characteristics within a gender-controlled set, then it should demonstrate a deviation lower than these baseline values.

\begin{table}[t!]
\caption{Consistency evaluation.}
\centering
\begin{adjustbox}{width=0.35\textwidth}
\begin{tabular}{ccccccc}
\hline\hline
\multirow{2}{*}{} & \multicolumn{2}{c}{Const(obj)$\uparrow$} & \multicolumn{2}{c}{Const(subj)$\uparrow$}\\
\cmidrule(lr){2-3} \cmidrule(lr){4-5}
 & Voiced & Silent & Voiced & Silent\\
\hline
GT & - & - & 3.80 & - \\
\hline
Base & 0.5611 & 0.5626 & 3.32 & 3.24\\ 
Flatten & \textbf{0.5631} & \textbf{0.5657} & 3.36 & 3.29\\ 
\hline\hline
\end{tabular}
\end{adjustbox}
\label{table:result2}
\vspace{-2mm}
\end{table}

\subsection{Results}
All results are averaged across both gender-wise outcomes and over all 10 trials. The \textit{Base} model refers to the proposed framework without the pitch-flattening module, while the \textit{Flatten} model includes the pitch-flattening module.

\subsubsection{Intelligibility evaluation}
Table~\ref{table:result1} shows the results of the intelligibility evaluation.
The overall error rate for the voiced test set is lower than that of the silent test set, consistent with the findings of previous research~\cite{gaddy2020digital}.
The \textit{Base} model shows a lower error rate on the voiced test set, however, the \textit{Flatten} model performs better on the silent test set ($p<0.05$ in paired t-tests). This suggest that the proposed pitch-flattening method is particularly effective for silent EMG signals, which is our primary target scenario, as we expected.

\subsubsection{Consistency evaluation}
Table~\ref{table:result2} shows the results of the consistency evaluation. In objective consistency, both the \textit{Base} and \textit{Flatten} model demonstrate better speaker embedding similarity than the random speaker similarity, which has a value of 0.5521 ($p<0.05$ in paired t-tests). Note that the random speaker similarity we calculated was based on the average results of the controlled gender set, which is a more stringent condition. This suggest that our model is capable of extracting information related to the target speaker's voice characteristics, beyond simple gender information from the face.
Additionally, the $Flatten$ model shows better speaker embedding similarity than the $Base$ model in both the voiced and silent test sets ($p<0.05$ in paired t-tests).
In subjective evaluation, the performance of two models shows no significance in either test set ($p>0.05$). 

\subsubsection{Pitch deviations evaluation}
Table~\ref{table:result3} presents the results of the pitch deviations evaluation. For \textit{Local F0 deviations}, $Flatten$ model outperforms the $Base$ model ($p<0.05$ in paired t-tests).
% It is worth noting that in the $Flatten$ model, the pitch-flattening module is applied not only before the content encoder in the EMG-based contents estimation network, but also before the content encoder in the face-based voice conversion network (see Figure~\ref{fig:figure1}).
% If local $F0$ deviation information is present in content embedding $c$, the frame-wise $F0$ decoder ($FF$) will extract this information directly from $c$, rather than estimate it using the content and speaker information.
% Then, the mismatch in $F0$ information within the EMG-based content embedding $c_{emg}$ could lead to performance degradation on the silent EMG test set.
% However, the pitch-flattening module in the $Flatten$ model forces the $FF$ to focus more on estimating $F0$ information, since the pitch-flattening module removes any pitch deviation information from the content embedding.
Therefore as we hypothesized in Section~\ref{sec:method1-4}, the proposed pitch-flattening module enhances our model's ability to estimate content-related pitch information.
% during the inference phase, the $Flatten$ model may produce more plausible frame-wise $F0$ values than the $Base$ model.

Additionally for the \textit{Global F0} evaluation, the $Flatten$ model outperforms the $Base$ model ($p<0.05$ in paired t-tests).
Similar to objective consistency evaluation, when reporting results by gender, the proposed $Flatten$ model not only shows better performance than the $Base$ model but also exhibits less error than the random deviation for each gender (29.18~Hz for males and 37.50~Hz for females).
This also supports that the model is not merely utilizing gender information for global pitch estimation but is also leveraging additional information related to the association between facial images and the voice characteristics of the target speaker.

\section{Discussion}
Through evaluations of intelligibility, consistency, and pitch deviations, the proposed multi-speaker silent EMG-to-speech generation framework demonstrates not only the ability to generate comprehensible speech, but also the capability to align the voice characteristics of the target speaker using only a facial image.
Additinally, the proposed pitch-flattening module proves effective in enhancing EMG based content embedding, particularly for silent EMG signals.

It is important to note that, the EMG dataset used in this study contains data from a single speaker, where content-related local pitch information is less varied compared to a multi-speaker dataset.
This suggests that the model may implicitly learn this information with or without the pitch-flattening module. Therefore, we anticipate that our proposed pitch-flattening module would yield greater performance gains on a multi-speaker EMG dataset, which would likely contain more diverse content-related local pitch information.

As mentioned in Section~\ref{sec:method}, all networks in our framework were trained from scratch using Gaddy's or the LRS3 dataset, except for the content encoder, which utilized pretrained ContentVec embeddings.
This means that the vocoding component, which is typically pretrained on clean dataset to produce higher-quality speech as used in previous research \cite{gaddy2020digital, scheck2023multi}, was also trained on the LRS3 dataset, which contains relatively lower-quality data due to its real-world settings.
We expect that employing conventional pretrained vocoders like \cite{kong2020hifi, lee2022bigvgan}, could enhance the generative quality of our proposed method, or that recent advancements in diffusion-based post-processing \cite{liu2022diffsinger} could also be applied.

Additionally, the pitch-flattening module we used does not benefit from recent advancements of pitch controlling techniques. It is a conventional signal processing method, yet the methodology we proposed has proven effective for EMG-based speech generation. However, recent research related to pitch-controlled speech generation~\cite{choi2021neural, choinansy++} might offer more stable pitch-flattening performance.

The scope of this work is limited to presenting a multi-speaker speech generation framework that operates without any audible inputs, and introduces a method using the pitch-flattening module. We leave further exploration of these enhancements as future work.

\begin{table}[t!]
\caption{Pitch deviations evaluation. `m' and `fm' refer to male and female, respectively, while `dv' represents deviations.}
\centering
\begin{adjustbox}{width=0.45\textwidth}
\begin{tabular}{ccccccc}
\hline\hline
\multirow{2}{*}{} & \multicolumn{2}{c}{Local $F0$ dv$\downarrow$} & \multicolumn{2}{c}{Global $F0$$\downarrow$}\\
\cmidrule(lr){2-3} \cmidrule(lr){4-5}
 & Voiced & Silent & Voiced (m/fm) & Silent (m/fm)\\
\hline
Base & 20.52 & 18.74 & 26.04/33.29 & 28.91/36.45 \\ 
Flatten & \textbf{17.39} & \textbf{17.93} & \textbf{24.98}/\textbf{31.56} & \textbf{24.80}/\textbf{30.83}\\ 
\hline\hline
\end{tabular}
\end{adjustbox}
\label{table:result3}
\vspace{-2mm}
\end{table}

\section{Conclusion}
In this work, we present a novel framework for generating multi-speaker speech using only facial inputs, a concept that has not been explored before. We utilized silent electromyography (EMG) signals to capture linguistic content and facial images to align with the vocal identity of the target speaker.
Additionally, we introduced a pitch-flattening method that enhances the EMG-based content embedding ability.
Our extensive analysis, including a proposed metric related to content-related local pitch deviation, demonstrates that our proposed framework not only achieves comprehensive speech quality but also effectively aligns with the voice characteristics of the target speaker without any audible inputs. We hope this research will pave the way for new possibilities for individuals with voice impairments.

\section{Acknowledgment}
This work was partly supported by Institute of Information \& communications Technology Planning \& Evaluation (IITP) grant funded by the Korea government(MSIT) [No. RS-2022-II220320, 2022-0-00320, 40\%], [No. RS-2022-II220641, 50\%], [No.RS-2021-II211343, Artificial Intelligence Graduate School Program (Seoul National University), 5\%], and [No.RS-2021-II212068, Artificial Intelligence Innovation Hub (Artificial Intelligence Institute, Seoul National University), 5\%].

% \section*{References}

\bibliographystyle{IEEEtran}
\bibliography{IEEEexample}

\end{document}